\documentclass[12pt]{article}
\newcommand{\bq}{\begin{equation}}
\newcommand{\eq}{\end{equation}}
\newcommand{\ba}{\begin{eqnarray}}
\newcommand{\ea}{\end{eqnarray}}

\newcommand{\nl }{ \nonumber  }

\newcommand{\p}{\partial}
\newcommand{\h}{\hspace{.5cm}}
\newcommand{\s}{\sigma}

\begin{document}
\vspace*{1.cm}
\begin{center}
{\bf SPIN CHAIN FROM MEMBRANE AND THE NEUMANN-ROSOCHATIUS
INTEGRABLE SYSTEM

\vspace*{1cm} P. Bozhilov}\footnote{e-mail: bozhilov@inrne.bas.bg}

\ \\
\textit{Institute for Nuclear Research and Nuclear Energy,\\
Bulgarian Academy of Sciences, \\ 1784 Sofia, Bulgaria}

\end{center}
\vspace*{.5cm}

We find membrane configurations in $AdS_4\times S^7$, which
correspond to the continuous limit of the $SU(2)$ integrable spin
chain, considered as a limit of the $SU(3)$ spin chain, arising in
$\mathcal{N}=4$ SYM in four dimensions, dual to strings in
$AdS_5\times S^5$. We also discuss the relationship with the
Neumann-Rosochatius integrable system at the level of Lagrangians,
comparing the string and membrane cases.



\vspace*{.5cm} {\bf Keywords:} M-theory, integrable systems,
AdS-CFT duality.

\section{Introduction}
One of the predictions of AdS/CFT duality is that the string
theory on $AdS_5\times S^5$ should be dual to $\mathcal{N}=4$ SYM
theory in four dimensions \cite{M97}, \cite{GKP98}, \cite{W98}.
The spectrum of the string states and of the operators in SYM
should be the same. The first checks of this conjecture {\it
beyond} the supergravity approximation revealed that there exist
string configurations, which in the semiclassical limit are
related to the anomalous dimensions of certain gauge invariant
operators in the planar SYM \cite{BMN02}, \cite{GKP02}. On the
field theory side, it was found that the corresponding dilatation
operator is connected to the Hamiltonian of integrable Heisenberg
spin chain \cite{MZ}. On the other hand, it was established in
\cite{K0311} that there is agreement at the level of actions
between the continuous limit of the $SU(2)$ spin chain arising in
$\mathcal{N}=4$ SYM theory and a certain limit of the string
action in $AdS_5\times S^5$ background. Shortly after, it was
shown that such equivalence also holds for the $SU(3)$ and $SL(2)$
cases \cite{HL04}, \cite{BCMS0409}\footnote{See also
\cite{AM0311,AM0402,AM0404, AM0409} and
\cite{KMMZ0402,KRT0403,DR0403,SJrT0404,KM0406,KT0406}.}.
Information about the latest developments on the subject can be
found for example in \cite{SJr0704} and references therein. Let us
also point out the work \cite{A0610}, where the connection between
the worldvolume theory of membranes living on flat space and
integrable quantum spin chains has been explored.

Here, we are interested in answering the question: is it possible
to reproduce this type of string/spin chain correspondence from
membranes on eleven dimensional curved backgrounds? It turns out
that the answer is positive at least for the case of M2-branes on
$AdS_4\times S^7$, as we will show later on. More precisely, we
will find that the action for the continuous limit of $SU(2)$
integrable spin chain, considered as a certain limit of the
$SU(3)$ chain, can be obtained from particular membrane
configurations.

This investigation is motivated by the following reasons. First of
all, it will shed light on some properties of the conjectured
duality between membranes on $AdS_4\times S^7$ and three
dimensional CFT \cite{M97}. In particular, our results will give a
support for the existence of an integrable sector on the field
theory side. On the other hand, strings on $AdS_5\times S^5$ and
membranes on $AdS_4\times S^7$ are dual to {\it different} gauge
theories. Hence, there should exist {\it common} integrable
sectors in the corresponding field theories.

The plan of the paper is as follows. In section 2, we reconsider
the $SU(3)\to SU(2)$ string case in diagonal worldsheet gauge.
Section 3 is devoted to membranes on $AdS_4\times S^7$. In section
4, we discuss the relationship with the Neumann-Rosochatius
integrable system at the level of Lagrangians, comparing the
string and membrane cases.

\setcounter{equation}{0}
\section{Strings on $AdS_5\times S^5$}
It is known that the ferromagnetic integrable $SU(3)$ spin chain
provides the one-loop anomalous dimension of single trace
operators involving the three complex scalars of $\mathcal{N}=4$
SYM. The nonlinear sigma-model, describing the continuum limit of
the $SU(3)$ spin chain, corresponds to strings moving with large
angular momentum on the five-sphere in $AdS_5\times S^5$
\cite{HL04}.

In order to have more close analogy with the membrane case
considered in the next section, we will reproduce the relevant
string action in the framework of diagonal worldsheet gauge. In
this gauge, the Polyakov action and constraints are given by
\ba\label{gfsa} &&S_{S}=\int d^{2}\xi \mathcal{L}_{S}= \int
d^{2}\xi\frac{1}{4\lambda^0}\Bigl[G_{00}-\left(2\lambda^0T\right)^2
G_{11}\Bigr],
\\ \label{gf00} &&G_{00}+\left(2\lambda^0T\right)^2 G_{11}=0,
\\ \label{gf01} &&G_{01}=0,\ea where
\ba\label{sim} &&G_{mn}= g_{MN}\p_m X^M\p_n X^N,\\ \nl
&&\left[\p_m=\p/\p\xi^m, \h m = (0,1),\h
(\xi^0,\xi^1)=(\tau,\sigma),\h M = (0,1,\ldots,9)\right],\ea is
the induced metric and $\lambda^0$ is Lagrange multiplier. The
commonly used conformal gauge corresponds to $2\lambda^0T=1$.

We choose to embed the string in $AdS_5\times S^5$ as follows
\ba\nl &&Z_{s}=R\mbox{r}_s(\xi^m)e^{i\phi_s(\xi^m)},\h
s=(0,1,2),\h
\eta^{rs} \mbox{r}_r\mbox{r}_s+1=0,\h \eta^{rs}=(-1,1,1), \\
\label{gse} &&W_{i}=Rr_i(\xi^m)e^{i\varphi_i(\xi^m)},\h
i=(1,2,3),\h \delta_{ij} r_ir_j-1=0,\ea where $\phi_s$ and
$\varphi_i$ are the isometric coordinates on which the metric of
$AdS_5$ and $S^5$ respectively does not depend. The embedding
coordinates $Z_{s}$, $W_{i}$ are related to the ones in
(\ref{sim}) by the equalities ($\phi_0=t$ is the time coordinate
on $AdS_5$)\footnote{Of course, other parameterizations of
$AdS_5\times S^5$, which ensure that the embedding constraints
$\eta^{rs} \mbox{r}_r\mbox{r}_s+1=0$ and $\delta_{ij} r_ir_j-1=0$
are satisfied identically, are also possible.}
\ba\nl &&Z_0= R\cosh\rho e^{i\phi_0},\\
\nl &&Z_1= R\sinh\rho\sin\theta e^{i\phi_1},\\ \nl &&Z_2=
R\sinh\rho\cos\theta e^{i\phi_2},\\ \nl &&W_1= R\sin\gamma\cos\psi
e^{i\varphi_1},\\ \nl &&W_2= R\sin\gamma\sin\psi e^{i\varphi_2},\\
\nl &&W_3=R\cos\gamma e^{i\varphi_3}.\ea

For the embedding (\ref{gse}), the induced metric takes the form
\ba\label{fsim} &&G_{mn}=\eta^{rs}\p_{(m}Z_r\p_{n)}\bar{Z_s} +
\delta_{ij}\p_{(m}W_i\p_{n)}\bar{W_j}= \\ \nl
&&R^2\left[\sum_{r,s=0}^{2}\eta^{rs}\left(\p_m\mbox{r}_r\p_n\mbox{r}_s
+ \mbox{r}_r^2\p_m\phi_r\p_n\phi_s\right) +
\sum_{i=1}^{3}\left(\p_mr_i\p_nr_i +
r_i^2\p_m\varphi_i\p_n\varphi_i\right)\right],\ea where $(...)$ on
the first line means symmetrization. The expression (\ref{fsim})
for $G_{mn}$ must be used in (\ref{gfsa}), (\ref{gf00}) and
(\ref{gf01}). Correspondingly, the string Lagrangian will be
\ba\label{fssl} \mathcal{L}= \mathcal{L}_{S}+ \Lambda_A(\eta^{rs}
\mbox{r}_r\mbox{r}_s+1) + \Lambda_S(\delta_{ij} r_ir_j-1),\ea
where $\Lambda_A$ and $\Lambda_S$ are Lagrange multipliers.

Here, we are interested in the following particular case of the
string embedding (\ref{gse}) \ba\nl Z_0=R e^{i\kappa\tau},\h
Z_1=Z_2=0,\ea which implies \ba\nl \mbox{r}_0=1,\h
\mbox{r}_1=\mbox{r}_2=0;\h \phi_0=t=\kappa\tau,\h \kappa=const.\ea
For this ansatz, $G_{mn}$ reduces to \ba\nl G_{mn}=
R^2\left[\sum_{i=1}^{3}\left(\p_mr_i\p_nr_i +
r_i^2\p_m\varphi_i\p_n\varphi_i\right)-\delta_{m}^{0}\delta_{n}^{0}\kappa^2\right].\ea
We now introduce new coordinates according to the rule \ba\nl
(\varphi_1,\varphi_2,\varphi_3)=
(\kappa\tau+\alpha+\varphi,\kappa\tau+\alpha-\varphi,\kappa\tau+\alpha+\phi)\ea
and take the limit $\kappa\to\infty$, $\p_0\to 0$, $\kappa\p_0$ -
finite. The result is \ba\nl &&G_{00}= 2R^2\kappa\left[\p_0\alpha
+ (r_1^2-r_2^2)\p_0\varphi+r_3^2\p_0\phi\right],
\\ \nl &&G_{11}= R^2\left\{\sum_{i=1}^{3}\left(\p_1r_i\right)^2 +(r_1^2+r_2^2)(\p_1\varphi)^2
+r_3^2(\p_1\phi)^2\right. \\ \nl &&+\left.(\p_1\alpha)^2+
2\p_1\alpha\left[(r_1^2-r_2^2)\p_1\varphi+r_3^2\p_1\phi\right]\right\},
\\ \nl &&G_{01}= R^2\kappa\left[\p_1\alpha +
(r_1^2-r_2^2)\p_1\varphi+r_3^2\p_1\phi\right].\ea In order to
eliminate $\p_1\alpha$, we use the constraint $G_{01}=0$ and
obtain the following two alternative expressions for the string
action after introducing the variable $t=\kappa\tau$ \ba\nl
S&=&\int d\tau d\sigma \mathcal{L}_{SC}=
\frac{R^2\kappa}{2\lambda^0}\int dt d\s\left[\p_t\alpha +
(r_1^2-r_2^2)\p_t\varphi + r_3^2\p_t\phi\right]\\ \nl
&-&\frac{\lambda^0(TR)^2}{\kappa}\int dt
d\s\left\{\sum_{i=1}^{3}\left(\p_1r_i\right)^2
+\left[(r_1^2+r_2^2)-(r_1^2-r_2^2)^2\right](\p_1\varphi)^2
+(r_1^2+r_2^2)r_3^2(\p_1\phi)^2\right.
\\ \nl
&-&\left. 2(r_1^2-r_2^2)r_3^2\p_1\varphi\p_1\phi\right\}+
\frac{1}{\kappa}\int dt
d\s\Lambda_S\left(\sum_{i=1}^{3}r_i^2-1\right)
\\ \label{nstra} &=&
\frac{R^2\kappa}{2\lambda^0}\int dt d\s\left[\p_0\alpha +
(r_1^2-r_2^2)\p_0\varphi +
r_3^2\p_0\phi\right]-\frac{\lambda^0(TR)^2}{\kappa}\int dt
d\s\left\{\sum_{i=1}^{3}\left(\p_1r_i\right)^2
\right.\\
\nl &+&\left.\frac{4r_1^2r_2^2}{r_1^2+r_2^2}(\p_1\varphi)^2
+(r_1^2+r_2^2)r_3^2\left[\left(\frac{r_1^2-r_2^2}{r_1^2+r_2^2}\right)\p_1\varphi-\p_1\phi\right]^2\right\}
\\ \nl &+&\frac{1}{\kappa}\int dt
d\s\Lambda_S\left(\sum_{i=1}^{3}r_i^2-1\right).\ea The momentum
$P_{\alpha}$ conjugated to $\alpha$ should be identified with the
total angular momentum of the string $J$ \ba\nl
P_{\alpha}=\frac{\pi R^2\kappa}{\lambda^0}\equiv J.\ea Then the
coefficients in the action become \ba\nl
\frac{R^2\kappa}{2\lambda^0}=\frac{J}{2\pi},\h
\frac{\lambda^0(TR)^2}{\kappa}=\frac{\lambda}{4\pi J},\ea where we
have used the relation $TR^2=\sqrt{\lambda}/2\pi$ between the
string tension $T$ and the 't Hooft coupling $\lambda$.

If we parameterize the two-sphere in the following way \ba\nl
r_1=\cos\psi\cos\theta,\h r_2=\sin\psi\cos\theta,\h
r_3=\sin\theta,\ea (\ref{nstra}) reduces to \ba\nl S&=&
\frac{J}{2\pi}\int dt d\s\left[\p_t\alpha +
\cos^2\theta\cos(2\psi)\p_t\varphi + \sin^2\theta\p_t\phi\right]\\
\nl &-&\frac{\lambda}{4\pi J}\int dt d\s\left\{\left(\p_\s
\theta\right)^2
+\cos^2\theta\left[\left(\p_\s\psi\right)^2+\sin^2(2\psi)(\p_\s\varphi)^2\right]\right.\\
\label{su3sc}
&+&\left.\frac{1}{4}\sin^2(2\theta)\left[\cos(2\psi)\p_\s\varphi
-\p_\s\phi\right]^2 \right\}.\ea This is the string action
corresponding to the thermodynamic limit of $SU(3)$ spin chain
after the identification $J\equiv L$ is made, where $L$ is the
length of the chain \cite{HL04}.

The particular case of $SU(2)$ spin chain corresponds to $r_3=0$
in (\ref{nstra}) or $\theta=0$ in (\ref{su3sc}). In order to make
connection with the membrane case, let us fix
$r_3^2=\varepsilon^2$ and take the limit $\varepsilon^2\to 0$ in
(\ref{nstra}). Neglecting the higher order terms, one obtains
\ba\label{rsa} S&=& \frac{R^2\kappa}{2\lambda^0}\int d\tau
d\s\left[\p_0\alpha + (r_1^2-r_2^2)\p_0\varphi\right]\\
\nl &-&\lambda^0(TR)^2\int d\tau
d\s\left\{\sum_{a=1}^{2}\left(\p_1r_a\right)^2
+\left[(r_1^2+r_2^2)-(r_1^2-r_2^2)^2\right](\p_1\varphi)^2\right\}\\
\nl &+& \int d\tau
d\s\Lambda_S\left[\sum_{a=1}^{2}r_a^2-(1-\varepsilon^2)\right].\ea
According to the constraint in the above action, the coordinates
$r_1$, $r_2$ must lie on a circle with radius
$(1-\varepsilon^2)^{1/2}$. To satisfy this constraint identically,
we choose \ba\nl r_1=(1-\varepsilon^2)^{1/2}\cos\psi,\h
r_2=(1-\varepsilon^2)^{1/2}\sin\psi,\ea and receive
\ba\label{saeps} S/(1-\varepsilon^2)&=&
\frac{R^2\kappa}{2\lambda^0}\int dt d\s\left[\p_t\tilde{\alpha} +
\cos(2\psi)\p_t\varphi\right]\\
\nl &-&\frac{\lambda^0(TR)^2}{\kappa}\int dt d\s
\left[\left(\p_\s\psi\right)^2+\sin^2(2\psi)(\p_\s\varphi)^2\right],\ea
where the new variable $\tilde{\alpha}$ has been introduced
through the equality \ba\nl
\alpha=(1-\varepsilon^2)\tilde{\alpha}.\ea Obviously, the right
hand side of (\ref{saeps}) coincides with the string action
corresponding to the thermodynamic limit of the $SU(2)$ spin chain
action.

\setcounter{equation}{0}
\section{Membranes on $AdS_4\times S^7$}
Turning to the membrane case, let us first write down the gauge
fixed membrane action and constraints in diagonal worldvolume
gauge, we are going to work with: \ba\label{omagf} &&S_{M}=\int
d^{3}\xi \mathcal{L}_{M}= \int
d^{3}\xi\left\{\frac{1}{4\lambda^0}\Bigl[G_{00}-\left(2\lambda^0T_2\right)^2\det
G_{ij}\Bigr] + T_2 C_{012}\right\},
\\ \label{00gf} &&G_{00}+\left(2\lambda^0T_2\right)^2\det G_{ij}=0,
\\ \label{0igf} &&G_{0i}=0.\ea They {\it coincide} with the
frequently used gauge fixed Polyakov type action and constraints
after the identification $2\lambda^0T_2=L=const$, where
$\lambda^0$ is Lagrange multiplier and $T_2$ is the membrane
tension. In (\ref{omagf})-(\ref{0igf}), the fields induced on the
membrane worldvolume $G_{mn}$ and $C_{012}$ are given by
\ba\label{im} &&G_{mn}= g_{MN}\p_m X^M\p_n X^N,\h C_{012}=
c_{MNP}\p_{0}X^{M}\p_{1}X^{N}\p_{2}X^{P}, \\ \nl
&&\p_m=\p/\p\xi^m,\h m = (0,i) = (0,1,2),\\ \nl
&&(\xi^0,\xi^1,\xi^2)=(\tau,\sigma_1,\sigma_2),\h M =
(0,1,\ldots,10),\ea where $g_{MN}$ and $c_{MNP}$ are the
components of the target space metric and 3-form gauge field
respectively.

Searching for membrane configurations in $AdS_4\times S^7$ dual to
integrable spin chains, we should first eliminate the membrane
interaction with the background 3-form field on $AdS_4$, to ensure
more close analogy with the strings on $AdS_5\times S^5$. To make
our choice, let us write down the background. It can be
parameterized as follows \ba\nl
&&ds^2=(2l_p\mathcal{R})^2\left[-\cosh^2\rho
dt^2+d\rho^2+\sinh^2\rho\left(d\alpha^2+\sin^2\alpha
d\beta^2\right) + 4d\Omega_7^2\right], \\ \nl
&&c_{(3)}=(2l_p\mathcal{R})^3\sinh^3\rho\sin\alpha dt\wedge
d\alpha\wedge d\beta.\ea Since we want the membrane to have
nonzero conserved energy and spin on $AdS$, the choice for which
the interaction with the $c_{(3)}$ field disappears is\footnote{Of
course, we can fix the angle $\beta$ instead of $\alpha$. We
choose to fix $\alpha$ because $\beta$ is one of the isometry
coordinates in the initial $AdS_4$ space.}: \ba\nl
\alpha=\alpha_0=const.\ea The metric of the corresponding subspace
of $AdS_4$ is \ba\label{sub}
&&ds^2_{sub}=(2l_p\mathcal{R})^2\left(-\cosh^2\rho
dt^2+d\rho^2+\sinh^2\rho\sin^2\alpha_0 d\beta^2\right)=
\\ \nl &&(2l_p\mathcal{R})^2\left[-\cosh^2\rho
dt^2+d\rho^2+\sinh^2\rho d(\beta\sin\alpha_0)^2\right].\ea Hence,
the appropriate membrane embedding into (\ref{sub}) and $S^7$,
analogous to the string embedding in $AdS_5\times S^5$, is \ba\nl
&&Z_{\mu}=2l_p\mathcal{R}\mbox{r}_\mu(\xi^m)e^{i\phi_\mu(\xi^m)},\h
\mu=(0,1),\h \phi_{\mu}=(\phi_0,\phi_1)=(t,\beta\sin\alpha_0),\\
\label{gme}
&&\h\h\h\h\h\h\h\h\h\h \eta^{\mu\nu} \mbox{r}_\mu\mbox{r}_\nu+1=0,\h \eta^{\mu\nu}=(-1,1), \\
\nl &&W_{a}=4l_p\mathcal{R}r_a(\xi^m)e^{i\varphi_a(\xi^m)},\h
a=(1,2,3,4),\h \delta_{ab} r_ar_b-1=0.\ea For this embedding, the
induced metric is given by \ba\label{mim}
&&G_{mn}=\eta^{\mu\nu}\p_{(m}Z_\mu\p_{n)}\bar{Z_\nu} +
\delta_{ab}\p_{(m}W_a\p_{n)}\bar{W_b}= \\ \nl
&&(2l_p\mathcal{R})^2\left[\sum_{\mu,\nu=0}^{1}\eta^{\mu\nu}\left(\p_m\mbox{r}_\mu\p_n\mbox{r}_\nu
+ \mbox{r}_\mu^2\p_m\phi_\mu\p_n\phi_\nu\right) +
4\sum_{a=1}^{4}\left(\p_mr_a\p_nr_a +
r_a^2\p_m\varphi_a\p_n\varphi_a\right)\right].\ea We will use the
expression (\ref{mim}) for $G_{mn}$ in (\ref{omagf}), (\ref{00gf})
and (\ref{0igf}). Correspondingly, the membrane Lagrangian becomes
\ba\label{geml}
\mathcal{L}=\mathcal{L}_{M}+\Lambda_A(\eta^{\mu\nu}
\mbox{r}_\mu\mbox{r}_\nu+1)+\Lambda_S(\delta_{ab} r_ar_b-1).\ea

In this paper, we are interested in the following particular case
of the membrane embedding (\ref{gme}) \ba\label{rme}
Z_{0}=2l_p\mathcal{R}e^{i\kappa\tau},\h Z_1=0,\ea which implies
\ba\nl \mbox{r}_0=1,\h \mbox{r}_1=0,\h \phi_0=t=\kappa\tau.\ea For
this ansatz, the metric induced on the membrane worldvolume
simplifies to \ba\label{mimpc} G_{mn}=(4l_p\mathcal{R})^2\left[
\sum_{a=1}^{4}\left(\p_mr_a\p_nr_a +
r_a^2\p_m\varphi_a\p_n\varphi_a\right)-\delta_{m}^{0}\delta_{n}^{0}(\kappa/2)^2\right],\ea
and the membrane Lagrangian is given by \ba\label{pcml}
\mathcal{L}=\mathcal{L}_{M}+\Lambda_S\left(\sum_{a=1}^{4}r_a^2-1\right).\ea

Let us now introduce new coordinates by setting \ba\nl
(\varphi_1,\varphi_2,\varphi_3,\varphi_4)=
\left(\frac{\kappa}{2}\tau+\alpha+\varphi,\frac{\kappa}{2}\tau+\alpha-\varphi,
\frac{\kappa}{2}\tau+\alpha+\phi,
\frac{\kappa}{2}\tau+\alpha-\tilde{\phi}\right).\ea Then, the
induced metric (\ref{mimpc}) takes the form \ba\nl
G_{mn}=(4l_p\mathcal{R})^2\left\{\frac{\kappa}{2}\left(\delta_{m}^{0}\p_n\alpha
+ \delta_{n}^{0}\p_m\alpha\right) +
\frac{\kappa}{2}\delta_{m}^{0}\left[(r_1^2-r_2^2)\p_n\varphi + r_3^2\p_n\phi - r_4^2\p_n\tilde{\phi}\right]\right.\\
\nl +
\left.\frac{\kappa}{2}\delta_{n}^{0}\left[(r_1^2-r_2^2)\p_m\varphi
+ r_3^2\p_m\phi - r_4^2\p_m\tilde{\phi}\right]+
\sum_{a=1}^{4}\p_mr_a\p_nr_a+ \p_m\alpha\p_n\alpha \right.\\ \nl +
\left.(r_1^2+r_2^2)\p_m\varphi\p_n\varphi + r_3^2\p_m\phi\p_n\phi
+ r_4^2\p_m\tilde{\phi}\p_n\tilde{\phi}\right.\\ \nl +
\left.(r_1^2-r_2^2)(\p_m\alpha\p_n\varphi + \p_n\alpha\p_m\varphi)
+ r_3^2(\p_m\alpha\p_n\phi + \p_n\alpha\p_m\phi) -
r_4^2(\p_m\alpha\p_n\tilde{\phi} +
\p_n\alpha\p_m\tilde{\phi}\right\}.\ea

Our next step is to take the limit $\kappa\to\infty$, $\p_0\to 0$,
$\kappa\p_0$ - finite. In this limit, $G_{00}$ and $G_{0i}$
simplify to \ba\nl &&G_{00}=
(4l_p\mathcal{R})^2\kappa\left[\p_0\alpha +
(r_1^2-r_2^2)\p_0\varphi+r_3^2\p_0\phi-r_4^2\p_0\tilde{\phi}\right],
\\ \nl &&G_{0i}= (4l_p\mathcal{R})^2\frac{\kappa}{2}\left[\p_i\alpha +
(r_1^2-r_2^2)\p_i\varphi+r_3^2\p_i\phi-r_4^2\p_i\tilde{\phi}\right],\ea
while $G_{ij}$ do not change. With the aim to take into account
the constraints $G_{0i}=0$ and to eliminate simultaneously
$\p_i\alpha$ from the membrane Lagrangian, we replace \ba\nl
-\p_i\alpha=
(r_1^2-r_2^2)\p_i\varphi+r_3^2\p_i\phi-r_4^2\p_i\tilde{\phi}\ea
into $\det G_{ij}$ and obtain for (\ref{pcml}) the following
expression \ba\label{limml}
\mathcal{L}&=&\frac{(2l_p\mathcal{R})^2}{\lambda^0}\kappa\left(\p_0\alpha
+ \sum_{k=1}^{3}\nu_k\p_0\rho_k\right)\\
\nl &-&\lambda^0T_2^2(4l_p\mathcal{R})^4
\left\{\sum_{a<b=1}^{4}(\p_1r_a\p_2r_b-\p_2r_a\p_1r_b)^2 +
\sum_{a=1}^{4}\sum_{k=1}^{3}\mu_k(\p_1r_a\p_2\rho_k-\p_2r_a\p_1\rho_k)^2\right.
\\ \nl &-&\left.\sum_{a=1}^{4}\left(\p_1r_a\sum_{k=1}^{3}\nu_k\p_2\rho_k -
\p_2r_a\sum_{k=1}^{3}\nu_k\p_1\rho_k\right)^2 +
\sum_{k<n=1}^{3}\mu_k\mu_n(\p_1\rho_k\p_2\rho_n-\p_2\rho_k\p_1\rho_n)^2\right.\\
\nl
&-&\left.\sum_{k=1}^{3}\mu_k\left(\p_1\rho_k\sum_{n=1}^{3}\nu_n\p_2\rho_n
- \p_2\rho_k\sum_{n=1}^{3}\nu_n\p_1\rho_n\right)^2\right\}
+\Lambda_S\left(\sum_{a=1}^{4}r_a^2-1\right).\ea For simplifying
reason, we introduced in (\ref{limml}) the notations \ba\nl
&&(\mu_1,\mu_2,\mu_3)=(r_1^2+r_2^2,r_3^2,r_4^2),\\ \nl
&&(\nu_1,\nu_2,\nu_3)=(r_1^2-r_2^2,r_3^2,-r_4^2),\\ \nl
&&(\rho_1,\rho_2,\rho_3)=(\varphi,\phi,\tilde{\phi}) .\ea

Now, we are ready to face our main problem: how to reduce the
membrane Lagrangian (\ref{limml}) to the one corresponding to the
thermodynamic limit of spin chain, {\it without shrinking the
membrane to string}? We propose the following solution of this
task: \ba\nl && \alpha=\alpha(\tau,\s_1),\h r_1=r_1(\tau,\s_1),\h
r_2=r_2(\tau,\s_1),\\ \label{oursol}
&&r_3=r_3(\tau,\s_2)=a\sin[b\s_2+c(\tau)],\h
r_4=r_4(\tau,\s_2)=a\cos[b\s_2+c(\tau)],\\ \nl
&&\varphi=\varphi(\tau,\s_1),\h a,b,\phi,\tilde{\phi}=constants,\h
a^2<1. \ea The restrictions (\ref{oursol}) lead to \ba\label{rml}
\mathcal{L}&=&\frac{(2l_p\mathcal{R})^2}{\lambda^0}\kappa\left(\p_0\alpha
+ \nu_1\p_0\rho_1\right)\\
\nl &-&\lambda^0(abT_2)^2(4l_p\mathcal{R})^4
\left[\sum_{a=1}^{2}(\p_1r_a)^2 +
(\mu_1-\nu_1^2)(\p_1\rho_1)^2\right]
+\Lambda_S\left[\sum_{a=1}^{2}r_a^2-(1-a^2)\right]\\ \nl
&=&\frac{(2l_p\mathcal{R})^2}{\lambda^0}\kappa\left[\p_0\alpha
+ (r_1^2-r_2^2)\p_0\varphi\right]\\
\nl &-&\lambda^0(abT_2)^2(4l_p\mathcal{R})^4
\left\{\sum_{a=1}^{2}(\p_1r_a)^2 +
\left[(r_1^2+r_2^2)-(r_1^2-r_2^2)^2\right](\p_1\varphi)^2\right\}
\\ \nl &+&\Lambda_S\left[\sum_{a=1}^{2}r_a^2-(1-a^2)\right].\ea
The above membrane Lagrangian is fully analogous to the string
Lagrangian in (\ref{rsa}), obtained after fixing $r_3^2$ to
$\varepsilon^2\to 0$. Proceeding as in the string case, we
introduce the parametrization \ba\nl r_1=(1-a^2)^{1/2}\cos\psi,\h
r_2=(1-a^2)^{1/2}\sin\psi,\ea the new variable $\tilde{\alpha}$
\ba\nl \alpha=(1-a^2)\tilde{\alpha},\ea and take the limit $a^2\to
0$. Thus, we receive \ba\label{rfc}
\mathcal{L}/(1-a^2)&=&\frac{(2l_p\mathcal{R})^2}{\lambda^0}\kappa\left[\p_0\tilde{\alpha}
+\cos(2\psi)\p_0\varphi\right]\\ \nl
&-&\lambda^0(abT_2)^2(4l_p\mathcal{R})^4\left[(\p_1\psi)^2+\sin^2(2\psi)(\p_1\varphi)^2\right],\ea
which should be compared with (\ref{saeps}). As for the membrane
action corresponding to the above Lagrangian, it can be
represented in the form \ba\label{maa} S_M&=&
\frac{\mathcal{J}}{2\pi}\int dt d\s\left[\p_t\tilde{\alpha} +
\cos(2\psi)\p_t\varphi\right]\\
\nl &-&\frac{\tilde{\lambda}}{4\pi\mathcal{J}}\int dt d\s
\left[\left(\p_\s\psi\right)^2+\sin^2(2\psi)(\p_\s\varphi)^2\right],\ea
where $\mathcal{J}$ is the angular momentum conjugated to
$\tilde{\alpha}$, $t=\kappa\tau$ and \ba\nl
\tilde{\lambda}=2^{15}[\pi^2(1-a^2)abT_2]^2(l_p\mathcal{R})^6.\ea
Obviously, the action (\ref{maa}) corresponds to the thermodynamic
limit of $SU(2)$ integrable spin chain.

\setcounter{equation}{0}
\section{Relationship with the Neumann-Rosochatius integrable system}
Here, we are going to discuss the connection between the
thermodynamical limits of the integrable spin chains, viewed as
arising from strings on $AdS_5\times S^5$ and membranes on
$AdS_4\times S^7$, and the Neumann-Rosochatius integrable system
\cite{Neumann,Per,NR}, related to specific configurations of
strings \cite{N,NR1,AAF05,NR2} and membranes \cite{MNR}.

\subsection{The string case}
It is known that large class of classical string solutions in the
type IIB $AdS_5\times S^5$ background is related to the Neumann
and Neumann-Rosochatius integrable systems. It was found in
\cite{NR2} that, working in conformal gauge, the spiky strings
\cite{MK,SR,IK0705,MS0705,BR0706} and giant magnons \cite{HM}-
\cite{PS07}, \cite{BR0706} can be also accommodated by a version
of the Neumann-Rosochatius system. The appropriate string
embedding of the type (\ref{gse}) is\footnote{As in section 2, we
will use diagonal worldsheet gauge.} \ba\label{NRa} Z_0=R
e^{i\kappa\tau},\h Z_1=Z_2=0,\h
W_{i}=Rr_i(\xi)e^{i[\omega_i\tau+f_i(\xi)]},\h
\xi=\alpha\sigma+\beta\tau,\ea i.e. \ba\nl \mbox{r}_0=1,\h
\mbox{r}_1=\mbox{r}_2=0;\h \phi_0=t=\kappa\tau, \h r_i=r_i(\xi),\h
\varphi_i=\omega_i\tau+f_i(\xi).\ea Correspondingly, the induced
metric takes the form \ba\nl
&&G_{00}=R^2\left\{\sum_{i=1}^{3}\left[\beta^2(\p_\xi r_i)^2 +
r_i^2(\beta\p_\xi f_i+\omega_i)^2\right]-\kappa^2\right\},\\ \nl
&&G_{11}=R^2\alpha^2\sum_{i=1}^{3}\left[(\p_\xi r_i)^2 +
r_i^2(\p_\xi f_i)^2\right],\\ \nl
&&G_{01}=R^2\alpha\sum_{i=1}^{3}\left\{\beta\left[(\p_\xi r_i)^2 +
r_i^2(\p_\xi f_i)^2\right]+\omega_ir_i^2\p_\xi f_i\right\}.\ea On
the Neumann-Rosochatius ansatz (\ref{NRa}), the string Lagrangian
is given by \ba\nl \mathcal{L}_*&=&-\frac{R^2}{4\lambda^0}
\left\{\sum_{i=1}^{3}\left[(A^2-\beta^2)(\p_\xi r_i)^2 +
(A^2-\beta^2)r_i^2\left(\p_\xi
f_i-\frac{\beta\omega_i}{A^2-\beta^2}\right)^2\right.\right.\\
\label{NRsl} &-& \left.\left.\frac{A^2}{A^2-\beta^2}\omega_i^2
r_i^2\right] + \kappa^2\right\} +
\Lambda_S\left(\sum_{i=1}^{3}r_i^2-1\right),\ea where \ba\nl
A^2\equiv \left(2\lambda^0T\alpha\right)^2.\ea After integrating
the equations of motion for $f_i$ once and replacing back the
solution into (\ref{NRsl}), one arrives at \ba\nl
\mathcal{L_*}&=&-\frac{R^2}{4\lambda^0}
\left\{\sum_{i=1}^{3}\left[(A^2-\beta^2)(\p_\xi r_i)^2 +
\frac{C_i^2}{(A^2-\beta^2)r_i^2}\right.\right.\\
\nl &-& \left.\left.\frac{A^2}{A^2-\beta^2}\omega_i^2 r_i^2\right]
+ \kappa^2\right\} +
\Lambda_S\left(\sum_{i=1}^{3}r_i^2-1\right),\ea where $C_i$ are
arbitrary integration constants. Following \cite{NR2}, we change
the overall sign, the signs of the terms $\sim C_i^2$, discard the
constant term $\sim\kappa^2$, and obtain \ba\nl
\mathcal{L}_{NR}=\frac{R^2}{4\lambda^0}
\sum_{i=1}^{3}\left[(A^2-\beta^2)(\p_\xi r_i)^2 -
\frac{C_i^2}{(A^2-\beta^2)r_i^2}
-\frac{A^2}{A^2-\beta^2}\omega_i^2 r_i^2\right] +
\Lambda_S\left(\sum_{i=1}^{3}r_i^2-1\right),\ea which is
Lagrangian for the Neumann-Rosochatius integrable system.

Now, with the aim of comparison with the thermodynamic limit of
the $SU(3)$ spin chain, let us set \ba\nl
\omega_1=\omega_2=\omega_3=\kappa,\h
(f_1,f_2,f_3)=(\alpha+\varphi,\alpha-\varphi,\alpha+\phi),\ea in
(\ref{NRsl}) and in the constraint $G_{01}=0$. The result is
\ba\nl \mathcal{L_*}&=&\frac{R^2}{4\lambda^0} \left\{2\kappa\beta
\left[\p_\xi\alpha +
(r_1^2-r_2^2)\p_\xi\varphi+r_3^2\p_\xi\phi\right]\right.\\
\nl &-&\left.(A^2-\beta^2)\left\{\sum_{i=1}^{3}(\p_\xi r_i)^2 +
(\p_\xi\alpha)^2
+2\p_\xi\alpha\left[(r_1^2-r_2^2)\p_\xi\varphi+r_3^2\p_\xi\phi\right]\right.\right.
\\ \nl
&+&\left.\left.(r_1^2+r_2^2)(\p_\xi\varphi)^2+r_3^2(\p_\xi\phi)^2\right\}\right\}
+ \Lambda_S\left(\sum_{i=1}^{3}r_i^2-1\right),\ea \ba\nl
&&\kappa\left[\p_\xi\alpha +
(r_1^2-r_2^2)\p_\xi\varphi+r_3^2\p_\xi\phi\right]
+\beta\left\{\sum_{i=1}^{3}(\p_\xi r_i)^2 +
(\p_\xi\alpha)^2\right.\\ \nl
&&+\left.2\p_\xi\alpha\left[(r_1^2-r_2^2)\p_\xi\varphi+r_3^2\p_\xi\phi\right]
+(r_1^2+r_2^2)(\p_\xi\varphi)^2+r_3^2(\p_\xi\phi)^2\right\}=0.\ea
The next step is to take the limit \ba\nl \kappa\to\infty,\
\beta\to 0,\h \kappa\beta - \mbox{finite},\ea in which
$\mathcal{L_*}$ reduces to \ba\label{NRlim}
\mathcal{L}_{lim}&=&\frac{R^2\kappa\beta}{2\lambda^0}
\left[\p_\xi\alpha +
(r_1^2-r_2^2)\p_\xi\varphi+r_3^2\p_\xi\phi\right]\\
\nl
&-&\left.\lambda^0(TR\alpha)^2\left\{\sum_{i=1}^{3}\left(\p_\xi
r_i\right)^2
+\left[(r_1^2+r_2^2)-(r_1^2-r_2^2)^2\right](\p_\xi\varphi)^2
+(r_1^2+r_2^2)r_3^2(\p_\xi\phi)^2\right.\right.
\\ \nl
&-&\left. 2(r_1^2-r_2^2)r_3^2\p_\xi\varphi\p_\xi\phi\right\} +
\Lambda_S\left(\sum_{i=1}^{3}r_i^2-1\right).\ea

Now, we turn to the case of string configurations corresponding to
the thermodynamic limit of the $SU(3)$ integrable spin chain, and
consider the particular case when all variables
($\alpha,\varphi,\phi,r_i$) depend only on
$\xi=\alpha\sigma+\beta\tau$. Then, $\mathcal{L}_{SC}$ in
(\ref{nstra}) {\it coincides} with $\mathcal{L}_{lim}$. Let us
point out that the constraint $G_{01}=0$ now reads \ba\nl
\p_\xi\alpha + (r_1^2-r_2^2)\p_\xi\varphi+r_3^2\p_\xi\phi=0.\ea

\subsection{The membrane case}
The most general membrane embedding in $AdS_4\times S^7$ leading
to the Neumann-Rosochatius integrable system is \cite{MNR} \ba\nl
&&Z_{0}=2l_p\mathcal{R}e^{i\kappa\tau},\h Z_1=0,\h
W_{a}=4l_p\mathcal{R}r_a(\xi,\eta)e^{i\left[\omega_{a}\tau+g_a(\xi,\eta)\right]},\\
\nl &&\xi=\alpha\sigma_1+\beta\tau,\h
\eta=\gamma\sigma_2+\delta\tau,\h \alpha,\beta,\gamma,\delta =
constants,\ea for \ba\nl &&r_{1}=r_{1}(\xi),\h r_{2}=r_{2}(\xi),\h
\omega_3=\pm\omega_4=\omega,\\ \nl
&&r_3=r_3(\eta)=a\sin(b\eta+c),\h r_4=r_4(\eta)=a\cos(b\eta+c),\h
a<1, \\ \nl &&g_1=g_1(\xi),\h g_2=g_2(\xi),\h
a,b,c,g_3,g_4=constants,\h \delta=0.\ea The above ansatz reduces
the membrane Lagrangian $\mathcal{L}$ in (\ref{geml}) to \ba\nl
\mathcal{L}_*^M&=&-\frac{(2l_p\mathcal{R})^2}{\lambda^0}
\left\{\sum_{a=1}^{2}\left[(\tilde{A}^2-\beta^2)(\p_\xi r_a)^2 +
(\tilde{A}^2-\beta^2)r_a^2\left(\p_\xi
g_a-\frac{\beta\omega_a}{\tilde{A}^2-\beta^2}\right)^2\right.\right.\\
\label{NRml} &-&
\left.\left.\frac{\tilde{A}^2}{\tilde{A}^2-\beta^2}\omega_a^2
r_a^2\right] + (\kappa/2)^2-(a\omega)^2\right\} +
\Lambda_S\left[\sum_{a=1}^{2}r_a^2-(1-a^2)\right],\ea where \ba\nl
\tilde{A}^2\equiv
\left(8\lambda^0T_2l_p\mathcal{R}ab\alpha\gamma\right)^2.\ea As
shown in \cite{MNR}, $\mathcal{L}_*^M$ corresponds to the
following Neumann-Rosochatius type Lagrangian \ba\nl
\mathcal{L}_{NR}^M&=&\frac{(2l_p\mathcal{R})^2}{\lambda^0}
\sum_{a=1}^{2}\left[(\tilde{A}^2-\beta^2)(\p_\xi r_a)^2 -
\frac{C_a^2}{(\tilde{A}^2-\beta^2)r_a^2}
- \frac{\tilde{A}^2}{\tilde{A}^2-\beta^2}\omega_a^2 r_a^2\right] \\
\nl &+&\Lambda_S\left[\sum_{a=1}^{2}r_a^2-(1-a^2)\right],\h
C_a=constants.\ea

Turning to comparison with the thermodynamic limit of the $SU(2)$
spin chain, we set in (\ref{NRml}) and in the constraint
$G_{01}=0$\footnote{For the present case, the constraint
$G_{02}=0$ is satisfied identically, due to $\delta=0$.} \ba\nl
\omega_1=\omega_2=\omega_3=\omega_4=\kappa/2,\h
(g_1,g_2)=(\tilde{\alpha}+\varphi,\tilde{\alpha}-\varphi)\ea which
leads to \ba\nl
\mathcal{L}_*^M&=&\frac{(2l_p\mathcal{R})^2}{\lambda^0}
\left\{\kappa\beta \left[(1-a^2)\p_\xi\tilde{\alpha} +
(r_1^2-r_2^2)\p_\xi\varphi\right]\right.\\
\nl &-&\left.(\tilde{A}^2-\beta^2)\left[\sum_{a=1}^{2}(\p_\xi
r_a)^2 + (1-a^2)(\p_\xi\tilde{\alpha})^2
+2(r_1^2-r_2^2)\p_\xi\tilde{\alpha}\p_\xi\varphi\right.\right.
\\ \nl
&+&\left.\left.(r_1^2+r_2^2)(\p_\xi\varphi)^2\right]\right\} +
\Lambda_S\left[\sum_{a=1}^{2}r_a^2-(1-a^2)\right],\ea \ba\nl
&&\frac{\kappa}{2}\left[(1-a^2)\p_\xi\tilde{\alpha} +
(r_1^2-r_2^2)\p_\xi\varphi\right]
\\ \nl
&&+\beta\left[\sum_{a=1}^{2}(\p_\xi r_a)^2 +
(1-a^2)(\p_\xi\tilde{\alpha})^2
+2(r_1^2-r_2^2)\p_\xi\tilde{\alpha}\p_\xi\varphi+(r_1^2+r_2^2)(\p_\xi\varphi)^2\right]=0.\ea

As a next step, we take the limit \ba\nl \kappa\to\infty,\
\beta\to 0,\h \kappa\beta - \mbox{finite},\ea in which
$\mathcal{L}_*^M$ and $G_{01}=0$ reduce to \ba\label{MNRlim}
\mathcal{L}_{lim}^M&=&\frac{(2l_p\mathcal{R})^2}{\lambda^0}\kappa\beta
\left[(1-a^2)\p_\xi\tilde{\alpha} +
(r_1^2-r_2^2)\p_\xi\varphi\right]\\
\nl &-&\lambda^0(T_2ab\alpha\gamma)^2(4l_p\mathcal{R})^4
\left\{\sum_{a=1}^{2}(\p_\xi r_a)^2 +
(1-a^2)\left[1-\left(\frac{r_1^2-r_2^2}{1-a^2}\right)^2\right](\p_\xi\varphi)^2\right\}
\\ \nl &+&
\Lambda_S\left[\sum_{a=1}^{2}r_a^2-(1-a^2)\right],\ea \ba\nl
(1-a^2)\p_\xi\tilde{\alpha} + (r_1^2-r_2^2)\p_\xi\varphi=0.\ea
Parameterizing the circle \ba\nl \sum_{a=1}^{2}r_a^2-(1-a^2)=0\ea
by \ba\nl r_1=(1-a^2)^{1/2}\cos\psi,\h
r_2=(1-a^2)^{1/2}\sin\psi,\ea one obtains \ba\label{MNRlimp}
\mathcal{L}_{lim}^M/(1-a^2)&=&\frac{(2l_p\mathcal{R})^2}{\lambda^0}\kappa\beta
\left[(\p_\xi\tilde{\alpha} +
\cos(2\psi)\p_\xi\varphi\right]\\
\nl &-&\lambda^0(T_2ab\alpha\gamma)^2(4l_p\mathcal{R})^4
\left[(\p_\xi \psi)^2 + \sin^2(2\psi)(\p_\xi\varphi)^2\right].\ea

Now, we turn to the case of membrane configurations corresponding
to the thermodynamic limit of the $SU(2)$ integrable spin chain.
Let us suppose that all variables ($\tilde{\alpha},\varphi,\psi$)
depend on ($\tau,\sigma_1$) through the linear combination
$\xi=\alpha\sigma_1+\beta\tau$ only. Then, $\mathcal{L}$ in
(\ref{rfc}) {\it coincides} with $\mathcal{L}_{lim}^M$ in
(\ref{MNRlimp}) for $\gamma=1$.

\setcounter{equation}{0}
\section{Concluding remarks}
In this paper we were able to find membrane configurations in
$AdS_4\times S^7$ of the type (\ref{gme}), given by (\ref{rme})
and (\ref{oursol}), which in a certain limit reproduce the
continuous limit of the $SU(2)$ integrable spin chain, arising in
$\mathcal{N}=4$ SYM, dual to strings on $AdS_5\times S^5$ (see
(\ref{maa})).

Besides, we investigated the connection between the
thermodynamical limits of the $SU(3)$ and $SU(2)$ integrable spin
chains, viewed as arising from strings on $AdS_5\times S^5$ and
membranes on $AdS_4\times S^7$, and the Neumann-Rosochatius
integrable system, related to specific string and membrane
configurations, at the level of Lagrangians. The conclusion is
that we can act in the same way in the string and membrane cases.
Namely, in order to obtain identical Lagrangians, we must impose
restrictions on both sides of the correspondence. On the side of
Neumann-Rosochatius model, we have to restrict ourselves to the
case when all frequencies are equal and proportional to the
parameter $\kappa$, related to the string/membrane energy, then
take an appropriate limit. On the spin chain side, we must
consider the particular case, when the dependence on the
coordinates $(\tau,\sigma/\sigma_1)$ is only through their linear
combination $\xi$.

As we pointed out in the introduction, in a recent work, the
connection between membrane dynamics on flat space-time (without
and with fluxes) and integrable quantum spin chains has been
discussed \cite{A0610}. In that paper, the relationship between
the worldvolume theory of membranes and integrable quantum spin
chains has been explored. More precisely, the author was able to
show that the very general framework for translating Hamiltonian
quantum mechanical matrix models to quantum spin chains in the
large N limit, which was first spelled out in \cite{AR0409}, can
be applied to explore the dynamics of a large class of membrane
models. This framework can presumably be of interest to the
membrane configurations, we are studying in this paper. Indeed, it
would be very illuminating if the underlying connection between
the integrable systems that were discussed in \cite{A0610} and the
ones arising here can be clarified.

On the other hand, according to AdS-CFT correspondence, strings on
$AdS_5\times S^5$ and membranes on $AdS_4\times S^7$ are dual to
{\it different} gauge theories. Therefore, one is tempting to
conjecture that there should exist common integrable sectors on
the field theory side.

\vspace*{.5cm} {\bf Note added:} While writing this paper, we
learn about the work \cite{0705.2634} on the same subject. There,
a rotating probe membrane in $S^3$ inside $AdS_4\times S^7$
background of M-theory is studied. With (partial) gauge fixing, it
is shown that in the fast limit, the worldvolume of tensionless
membrane reduces to either the $XXX_{1/2}$ spin chain or the
two-dimensional $SU(2)$ Heisenberg spin model. After that, the
author introduces anisotropy and coupling to an external magnetic
field. The correspondence for higher dimensional (D)p-branes is
also considered.

\vspace*{.5cm}

{\bf Acknowledgments} \vspace*{.2cm}

This work is supported by NSFB grants $F-1412/04$ and
$VU-F-201/06$.


\end{document}